\documentclass[aps,prl,floatfix,twocolumn,showpacs,reprint]{revtex4-1}
\usepackage{amsfonts}
\usepackage{mathrsfs}
\usepackage{amsmath}
\usepackage{color}
\usepackage{graphicx}
\usepackage{bm}
\usepackage{amssymb}
\usepackage{xspace}
\usepackage{epstopdf}
\usepackage{dcolumn}
\usepackage{longtable}
\usepackage{ulem} 
\usepackage{multirow}
\usepackage[colorlinks=true, letterpaper=true, pdfstartview=FitV, linkcolor=red, citecolor=blue, urlcolor=red]{hyperref}
\usepackage[colorinlistoftodos,prependcaption,textsize=tiny]{todonotes}


\begin{document}

\title{ Nonlinear dynamics of  a spin-orbit-coupled Bose-Einstein condensate }

 \author{Yongping Zhang}
	 \email{yongping11@t.shu.edu.cn}
	\affiliation{Department of Physics, Shanghai University,  Shanghai 200444, China}

 \author{Zhiqian Gui}
\affiliation{Department of Physics, Shanghai University,  Shanghai 200444, China}

 \author{Yuanyuan Chen}
\affiliation{Department of Physics, Shanghai University,  Shanghai 200444, China}

 \begin{abstract}

Single-particle dynamics of a spin-orbit-coupled Bose-Einstein condensate  has recently been investigated in experiments that explore the physics of Landau-Zener tunneling and of the Zitterbewegung. In this paper, we study the influence of a nonlinearity due to interactions on these dynamics and show that the dispersion relation develops interesting loop structures.  The atoms can move along the nonlinear dispersion curve in the presence of a weak acceleration force and  we show that the loops lead to straightforward nonlinear Landau-Zener tunneling. However, we find that for the Zitterbewegung, induced by a sudden quench in the spin-orbit coupling parameters, the nonlinear dispersion is irrelevant.

\end{abstract}

\pacs{ 03.75.Mn, 71.70.Ej}




\maketitle

 Since the experimental implementation of artificial spin-orbit coupling in atomic Bose-Einstein condensates (BECs)~\cite{Lin}, a great deal of attention has been paid to the physics of spin-orbit-coupled neutral atoms (see, e.g.,~\cite{Goldman2014, JingZhang,Zhai,Xu2015,Li2015,Shizhong2015,Yongping2016} and reference therein).  For a two-component BEC, experimentally realizable one-dimensional spin-orbit coupling~\cite{Lin,Fu2011} gives rise to an energy-quasimomentum dispersion relation featuring two energy bands with a double well structure in lower branch. This novel dispersion behaviour greatly enriches the ground-state phase diagram~\cite{Wang2010,Ho2011,Wucongjun2011,Sinha2011,Hu2012,Li2012,Li2013,Hamner2014} and the collective excitation dynamics~\cite{Pan2012,Zheng2013,Khamehchi2014,Pan2015}, compared with the parabolic dispersion of ordinary atomic BECs. By control over experimental parameters, dynamics involving two spin-orbit-coupled bands can be excited and it has been shown experimentally that a quantum quench can result in intriguing Zitterbewegung oscillations~\cite{Qu}. Experimental confirmation also exists for identifying the transitions between two bands due to an acceleration as being of Landau-Zener type~\cite{Olson}. Both of these two experiments deal with many-body systems, in which the nonlinearity stemming from the atomic interactions cannot be neglected. Nevertheless, the experimental measurements can be completely explained by utilizing single-particle spin-orbit-coupled bands~\cite{Qu, Olson,Xiong, Llorente} and the role of nonlinearity in these dynamics remains to be explored. Very recently, the experimental group at MIT has pushed spin-orbit-coupled BECs into a new regime where the nonlinear coefficients are fully tunable~\cite{Ketterle}.

In this work, we theoretically study the dynamics of a spin-orbit-coupled BEC and explicitly consider the nonlinear effects.  Similar to the experiments in Refs.~\cite{Qu, Olson}, we study two different approaches to trigger nonlinear dynamics, one is by a weak acceleration force and the other is by a quench. We intuitively expect that the nonlinear dynamics follows the nonlinear energy-quasimomentum dispersion based on single-particle physics. However, our results demonstrate that depending on the way the dynamics is  initiated, the system responds distinctly different to the nonlinear dispersion.
 When the acceleration force is weak compared to the energy difference between the two spin-orbit-coupled bands, it drives the atoms to adiabatically move along the nonlinear spin-orbit-coupled bands. In addition, for certain nonlinear coefficient regimes, a loop-like structure appears in the nonlinear dispersion relation.  This is an entirely nonlinear effect without a linear analog. The loops break the adiabatic motion of the atoms, even while the acceleration force is weak, leading to nonlinear Landau-Zener transitions \cite{Wu,Liu,Chen} between two nonlinear spin-orbit-coupled bands. We propose to measure the sizes of the loop structures by taking advantage of nonlinear Landau-Zener transitions. We show that for the acceleration force induced dynamics the nonlinear dispersion relation is fully relevant. We next study quench dynamics induced by a sudden jump of the spin-orbit coupling parameters and show that the resulting dynamics has no connection to the nonlinear dispersion. The resulted dynamics shows periodic oscillation, whose frequency does not match with energy difference between the two nonlinear bands, which is distinctly different from the single-particle dynamics,
  where atoms oscillate between single-particle bands.
  In fact, the presence of a nonlinearity breaks the quantum superposition principle, such that nonlinear bands can not form a complete basis to support dynamics.
 Despite the presence of the nonlinearity in the system, the quench dynamics experiments in~\cite{Qu} follow single-particle bands, and we show that the experiments can therefore be effectively interpreted as single-particle systems.

\begin{figure*}[t]
\includegraphics[bb=17 17 278 63, width=1\textwidth]{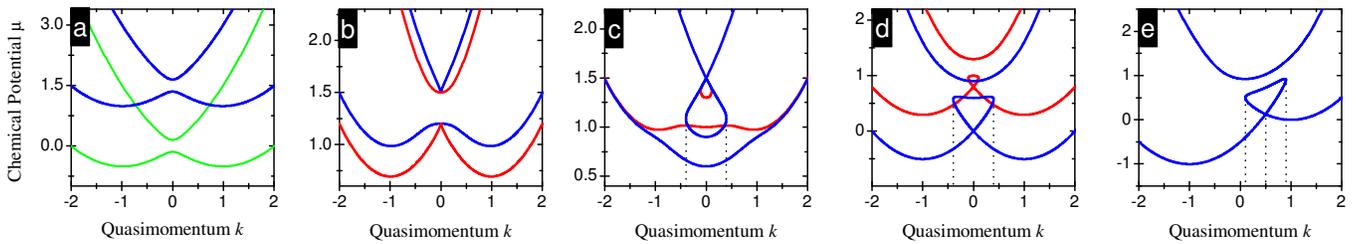}
\caption{ Nonlinear dispersion relation $\mu(k)$ for $g_{11}=g_{22}=g$,  $\Omega=0.3$ and $\gamma=1$.   (a): Comparison of the nonlinear dispersion with equal nonlinear coefficients $g_{12}=g=1.5$ (blue lines) with the linear dispersion $g_{12}=g=0$ (green lines). (b): Cusps can be seen to develop in the upper branch (blue lines) for $g=1.5, g_{12}=1.2$ and in the lower branch (red lines) for $g=1.2, g_{12}=1.5$. (c): Loops appear in the upper branch and their sizes change with different nonlinear coefficients. Red lines are for $g=1.5$ and $g_{12}=0.8$, and blue lines are for $g=1.5$ and $g_{12}=0$. (d): Loops exist in the lower branch, red lines are for  $g=0.8$ and $g_{12}=1.5$ and blue lines are for $g=0$ and $g_{12}=1.5$. In (a)-(d), $\delta=0$. (e): Loop structure with a finite detuning $\delta=-1$ for $g=0$ and $g_{12}=1.5$.  The vertical dashed lines in (c)-(e) label the locations of the edges and the center of the loop structures.
}
\label{Bands}
\end{figure*}

The dynamics of a spin-orbit-coupled Bose-Einstein condensate obeys  the dimensionless  Gross-Pitaevskii (GP) equation,
\begin{equation}
i\frac{\partial \Psi }{\partial t} = (  H_\text{soc}+ H_\text{non} [  \Psi ]  )  \Psi,
\label{GP}
\end{equation}
where the wave function is a spinor $\Psi=[\Psi_1,\Psi_2]^T$. $H_\text{soc}$ is the single-particle spin-orbit-coupled  Hamiltonian,
\begin{equation}
H_\text{soc}= -\frac{1}{2} \frac{\partial^2}{\partial x^2}  -i\gamma \frac{\partial }{\partial x} \sigma_z +   \frac{\delta}{2} \sigma_z +\frac{\Omega}{2} \sigma_x,
\end{equation}
and $ H_\text{non} $ accounts for the interactions,
\begin{equation}
 H_\text{non}[\Psi] =  \begin{pmatrix}    g_{11} |\Psi_1|^2 +g_{12} |\Psi_2|^2  &0 \\  0 &      g_{12} |\Psi_1|^2 +g_{22} |\Psi_2|^2  \end{pmatrix}.
\end{equation}
The one dimensional spin-orbit coupling is generated experimentally by applying two Raman beams that couple two pseudo-spin states \cite{Lin, Fu2011, Hamner2014, Pan2012, Khamehchi2014, Pan2015, Qu, Olson, Ketterle}.
 $\gamma$ characterizes the spin-orbit coupling strength  $\hbar k_\text{Ram}/m$ with $ k_\text{Ram}$ being wave number of Raman beams, $\delta$ is the detuning, and $\Omega$ is the Rabi frequency. In the dimensionless GP equation, the units of energy, length and time are  $\hbar^2 k_\text{Ram}^2/m$ , $1/k_\text{Ram}$ and $m/(\hbar k_\text{Ram}^2)$, respectively. Such scales lead to $\gamma =1$, however it can be varied to be $\gamma \le1$ by performing a high-frequency driving of the intensity of the Raman lasers~\cite{Jimenez,Yongping}. We therefore only consider $0< \gamma \le 1$, since the physics in the $\gamma <0$ regime is same as that in the regime of $\gamma>0$. This is because the GP equation possesses a spin rotation symmetry, i.e., $U=\sigma_x$, when also flipping the signs of $\gamma$ and $\delta$.
 The nonlinear coefficients $g_{ij}$ are proportional to the s-wave scattering lengths and atom numbers, and can be adjusted in experiments~\cite{Ketterle}.

Guided by the fact that single-particle dynamics follows from the single-particle energy dispersion relation, we first analyse the nonlinear dispersion relation of our system in order to understand the nonlinear dynamics. This can be done by
assuming that the solution of Eq.~(\ref{GP}) is a plane wave function, i.e., $\Psi=\exp(ikx-i\mu t) \Phi$, with $k$ being quasimomentum in units of $k_{Ram}$, and $\mu$ being the chemical potential. $\Phi=[\Phi_1,\Phi_2]^T$ then admits a time-independent GP equation,
\begin{equation}
\mu\Phi = \left( H_\text{soc}(k)  \Phi + H_\text{non}[\Phi] \right)    \Phi,
\label{TIGP}
\end{equation}
where the single-particle Hamiltonian in momentum space  can be written as,
$
H_\text{soc}(k) = k^2/2  + \gamma k  \sigma_z +   \delta \sigma_z /2+\Omega \sigma_x/2.
$
Meanwhile, $\Phi$ satisfies the renormalization condition $|\Phi_1|^2+|\Phi_2|^2=1$.

The nonlinear dispersion relation $\mu(k)$, i.e., chemical potential as a function of the quasimomentum $k$, can be numerically  calculated from Eq.~(\ref{TIGP}) for different values of the nonlinear coefficients. For simplicity, in the following, we consider $g_{11}=g_{22}=g$ and show typical results in Fig.~\ref{Bands}. Since we consider two atomic components, the dispersion relation possesses two branches. If all nonlinear coefficients are equal, i.e.~$g_{12}=g$, we find that the nonlinear dispersion is just an overall upward displacement of its linear counterpart [see Fig.~\ref{Bands}(a)]. If the inter-species interaction is not equal to the intra-species interaction, i.e.~$g_{12} \ne  g$, the nonlinear dispersion relation develops several interesting features: (1) at certain values of the nonlinear coefficients a cusp appears in upper or lower band [see Fig.~\ref{Bands}(b)]; (2) loop structures can arise in the upper or lower branch [see  Figs.~\ref{Bands}(c) and \ref{Bands}(d)]. The sizes of the loops depend on the nonlinear coefficients; (3) moreover, the detuning can tilt the loops and shift their location in momentum space, as shown in Fig.~\ref{Bands}(e).

Spin-orbit-coupled BECs are not the only systems to support the existence of loop structures in the nonlinear dispersion relation. They have been investigated theoretically and experimentally in various nonlinear systems, ranging from BECs in optical lattices~\cite{Wu1, Diakonov, Mueller, Machholm, Seaman, Danshita, Lee, Koller}, in optical cavities~\cite{Venkatesh, Dong}, in Bragg gratings~\cite{Porsezian}, with Dirac cones~\cite{Zhu, Bomantara}, and exciton-polariton condensates~\cite{Chestnov} to degenerate Fermi gases~\cite{Gentaro,Yu}. A typical and simple system to study loop-features in nonlinear dispersion relations is a two-level avoided crossing~\cite{Wu, Liu}. When nonlinear effects overcome the gap size of the avoided crossing, some nonlinear solutions without linear analogues emerge and, interestingly, they can form  closed loops in the dispersion relation. It is interesting to note that in the spin-orbit-coupled BECs there is no global band gap opening. Loops locate inside a local energy gap, which is different from those in above mentioned systems where the position of loops is inside global band gaps. Similar loop structures in local gaps are shown in~\cite{Ye}.

In order to understand the underlying physics of the general results depicted in Fig.~\ref{Bands}, we therefore map our system onto a standard nonlinear two-level model~\cite{Wu, Liu}. By using the renormalization condition $  |\Phi_1|^2   + |\Phi_2|^2=1$, we re-write Eq.~(\ref{TIGP}) as, $\mu_\text{ntl} \Phi = H_\text{ntl}\Phi $, with $\mu_\text{ntl}=\mu - (g+g_{12})/2-k^2/2$, and the nonlinear two-level Hamiltonian $ H_\text{ntl}$ is,
\begin{equation}
 H_\text{ntl} = \left(\gamma k +\frac{\delta}{2} \right) \sigma_z + \frac{g_{12}-g}{2}\left( |\Phi_2|^2- |\Phi_1|^2   \right) \sigma_z  +\frac{\Omega}{2}\sigma_x,
 \label{NTL}
\end{equation}
where the $\sigma$ are the Pauli matrices. The linear parts (the first and third term) defines an avoided crossing in momentum space, and an energy gap is opened of size $\Omega$ at $k= -\delta/(2\gamma)$. The competition between the nonlinear (second term) and the linear parts is mainly represented by the non-commutation between the nonlinear term $(g_{12}-g) ( |\Phi_2|^2- |\Phi_1|^2   ) \sigma_z/2$  and the gap size term  $\Omega\sigma_x/2$. When the nonlinear term dominates over the gap size, $| g_{12} -g | > \Omega $, loop structures appear~\cite{Wu, Liu}. The critical condition $| g_{12} -g | = \Omega $ is special, as at this point a cusp [see Fig.~\ref{Bands}(b)] develops to prepare the dispersion relation for the  emerging loop structures. When $g_{12}-g>\Omega$, new nonlinear solutions without linear analogues bifurcate from the lower branch and extend upwards into energy gap since the effective nonlinearity is repulsive if $g_{12}-g >0$. These new solutions form a loop attached to the lower dispersion branch as seen in Fig.~\ref{Bands}(d). If $g_{12}-g <0$, the effective nonlinearity is attractive, and when
$g_{12}-g <- \Omega$, the loop structure bifurcates from the upper branch [see Fig.~\ref{Bands}(c)].

The center of the loops locates at $k= -\delta/(2\gamma)$, where the gap in the linear dispersion relation has a minimum. Therefore, the finite value of $\delta$ can displace the location of loops, as can be seen in Fig.~\ref{Bands}(e). Since the chemical potential $\mu$ is dressed by the quasimomentum as $  \mu= \mu_\text{ntl}+ (g+g_{12})/2+ k^2/2$, the deviation of the loop center from $k=0$ must be accompanied by a tilt of loop due to the asymmetry in $k^2/2$ with respect to $k\ne 0$.
The loop size can be calculated following the results in Ref.~\cite{Liu} and the region over which the loop extends in momentum space is given by $-k_{c} -\delta/(2\gamma) \le k \le k_{c} -\delta/(2\gamma)$
with $k_c=
 \left( |g_{12}-g|^\frac{2}{3} -\Omega^\frac{2}{3} \right)^\frac{3}{2} /( 2\gamma) $. The edges of loop are at $\pm k_c -\delta/(2\gamma)$ and are shown by vertical dashed lines in
Figs.~\ref{Bands}(c)-(e). The size of the loops depends on the parameters $|g_{12}-g|^\frac{2}{3} -\Omega^\frac{2}{3} $ and $\gamma$. Two loops of different sizes for different values of $|g_{12}-g|$ are shown in  Figs.~\ref{Bands}(c) and (d) .

\begin{figure}[t]
\includegraphics[bb=17 13 251 218,width=0.5\textwidth]{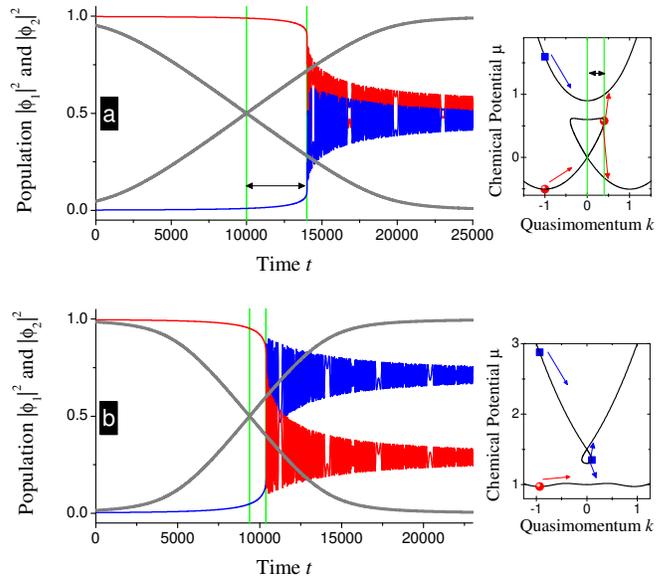}
\caption{ Nonlinear Landau-Zener dynamics in the presence of the loop structures. The acceleration force is given by $F=0.0001$, and the other parameters are $\Omega=0.3, \gamma=1$ and $\delta=0$. The right panels show the nonlinear dispersion relations and the left panels display the dynamics for different initial states. (a): Loop in the lower band. The initial state corresponding to the thick grey lines in the left plot is prepared in the upper band (square in the right plot) and the initial state for the thin lines is in the lower band (the lower circle in the right plot). $g_{11}=g_{22}=g=0$ and $g_{12}=1.5$. (b): Loop in the upper band. The initial state for the thick lines in the left plot corresponds to the circle in the right plot, and that for the thin lines corresponds the square in the right plot. $g_{11}=g_{22}=g=1.5$ and $g_{12}=0.8$.
}
\label{LandauZener}
\end{figure}

The existence of the loop structures leads to nontrivial dynamics.  To show this we study nonlinear Landau-Zener transitions by applying a very weak constant acceleration force $F$ to the atoms~\cite{Olson, Xiong, Llorente}. The BEC is initially prepared in the ground state of the system labeled by the lower red dot in Fig.~\ref{LandauZener}(a). The constant force accelerates atoms to move along the nonlinear dispersion [indicated by an arrow in Fig.~\ref{LandauZener}(a)], and
once the atoms arrive at the edge of the loop [indicated by a vertical line], there is no trajectory for them to go and they have no choice but to jump to the upper or lower bands. The finite possibility for this jump to bring the atom into the upper band constitutes a nonlinear Landau-Zener transition, no matter how weak the acceleration force is. Since the nonlinear Landau-Zener transition physically originates from loop structures, it does not have a linear analogue and is a fully nonlinear effect. We stimulate the dynamics by numerical integration of Eq.~(\ref{GP}) in momentum space, The acceleration force $F$ changes quasimomentum
as $k(t)= k_\text{ini}+Ft$ with $k_\text{ini}$ being initial quasimomentum~\cite{Aschcoft}.
The time evolution of the population in $\Phi_1$ and $\Phi_2$ for different initial states is shown in Fig.~\ref{LandauZener}(a), where thin lines correspond to the initial state in the lower nonlinear band and the initial state of the data shown by the thick grey lines is in the upper band [represented by a blue square].

The dynamics for these two different initial states can be seen to be quite different. Due to a finite energy gap and a weak acceleration force, atoms initially in the upper band smoothly move along the upper band and change their spin-compositions. At a certain time [indicated by a vertical line], the two components become population-balanced, which in the dispersion relation corresponds to the atoms arriving at $k=0$ in the upper band and the spin becoming unpolarized. Due to the existence of the loop structure, the atoms in the lower band undergo nonlinear Landau-Zener transition. After they arrive at the loop edge, the populations show fast oscillations. The oscillatory behavior exists even in the linear analogous systems. In linear systems, after Landau-Zener transition in the vicinity of the avoided crossing, atoms evolve in both upper and lower bands, the beating between two bands arises the oscillation. Furthermore, the oscillatory period changes since the energy difference between two bands changes for different quasimomentums. The single-particle phenomena of the transition probability and oscillatory behavior have been experimentally observed in spin-orbit-coupled BECs~\cite{Olson}. As shown by thin lines in the left panel of  Fig.~\ref{LandauZener}(a), before the atoms arrive at the loop edge, the spin-compositions evolve smoothly, and follow the dispersion relation. Different initial states respond differently to the same acceleration force, which can be used to experimentally measure the size of loop structures. The distance between the two vertical lines in the left panel exactly corresponds to the length between $k=0$ and the loop edge in the dispersion relation depicted in the right panel.  In experiments, the time evolution of the populations can be measured in a spin-resolved Stern-Gerlach time-of-flight setup.

We also numerically check the case where the loop structures appear in the upper band. Typical results are shown in Fig.~\ref{LandauZener}(b). Since the loop occurs in the upper band, atoms initially in the upper band can undergo nonlinear Landau-Zener transition, with the fast oscillations [thin lines in the left panel] indicating that the atoms have arrived at the loop edge in the dispersion. In this case the atoms initially in the lower band evolute smoothly. Similar to the case of loops in the lower band, one may compare the dynamics of different initial states to extract the information about the loop size.


The above dynamics of nonlinear Landau-Zener tunneling reveals that a nonlinear dispersion relation truly has a fundamental effect on nonlinear dynamics. This motivates us to study further nonlinear effects in quench dynamics, where a sudden jump of the Rabi frequency from $-\Omega$ to $\Omega$ occurs. Such a sign change can be  implemented in experiments by adjusting the relevant phases between the two Raman lasers.
The changing of the sign of $\Omega$ does not modify the nonlinear dispersion relation, as equation~(\ref{TIGP}) has the symmetry $\mathcal{P}_\Omega \sigma_z$ with $ \mathcal{P}_\Omega$ flipping the sign of $\Omega$, i.e., $\mathcal{P}_\Omega^\dagger \Omega \mathcal{P}_\Omega=- \Omega$. However, the eigenstates of systems with opposite signs of $\Omega$ are different.  Such a quench technique has, for example, already been used to explore the Zitterbewegung in  spin-orbit-coupled atoms~\cite{Qu}.

\begin{figure}[t]
\includegraphics[bb=16 13 241 215]{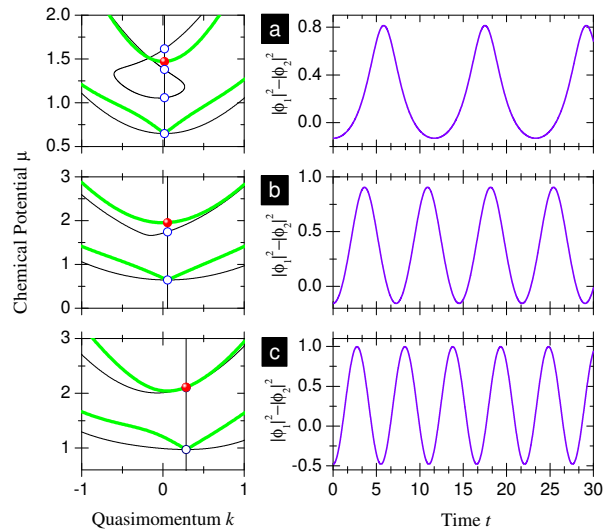}
\caption{ (Color Online) Nonlinear dynamics induced by the quenching of the Rabi frequency from $-\Omega$ to $\Omega$ for different values of the nonlinear coefficients and $\gamma=0.6$ and $\delta=0.2$. In the right hand side panels, the quench dynamics is shown by the evolution of the population difference $|\Phi_1|^2-|\Phi_2|^2$. In the left hand side panels, the nonlinear dispersion $\mu(k)$ (thin back lines) and the collective excitations $\omega(q)$ (thick green lines) are displayed together. Open circles are eigenvalues for a fixed quasimomentum, and solid circles are the excitations for a fixed quasimomentum of excitations. (a): Nonlinear dynamics in the present of a loop structure in the upper band. $\Omega=0.4$, $g_{11}=g_{22}=g=1.5$ and $g_{12}=0.2$. The values of the open circles are $\mu=(1.62, 1.38, 1.06, 0.65)$ from top to bottom. The value of solid circle is $\mu=1.47$. The oscillation period is $T=11.67$. (b) Nonlinear dynamics in the absence of loop structure. $ \Omega=1 $, $g=1.5$ and $g_{12}=0.8$. The open circles represent $\mu=(0.65,1.75)$, and the location of solid circle is at $\mu=1.95$. The oscillation period is $T=7.26$.  (c) Effective linear dynamics with the parameters $\Omega=1$ and $g=g_{12}=1.5$. The open and solid circles overlap, $\mu=(2.11,0.97) $ from top to bottom. The oscillation period is $T=5.52$, which exactly matches with the energy difference between the top and bottom circles.
}
\label{Quench}
\end{figure}

To study the quench dynamics in the presence of loops, we follow the experiment procedure described in~\cite{Qu} and prepare ground states in the lower band [indicated by the lowest dots in the left panels of Fig.~\ref{Quench}]. We then track the evolution of the population difference $|\Phi_1|^2-|\Phi_2|^2$ after changing from $-\Omega$ to $\Omega$. Since it is known from experiments  that the quench does not change the quasimomentum for a very long time~\cite{Qu}, we will only use a fixed quasimomentum in the following.  In the presence of loops, there are four bands for a fixed quasimomentum, which are represented by open dots in Fig.~\ref{Quench}(a). One may intuitively expect that the  dynamics following the quench would be a complicate oscillation between these four states, possessing multiple periods. However, very surprisedly, the population difference evolution shown in the right panel of  Fig.~\ref{Quench}(a) indicates that the dynamics is periodic but with only one frequency. Furthermore, this frequency does not match up with the energy differences between any two of the four bands.

This indicates that the loops are completely irrelevant to the quench dynamics and we compare these results to a parameters region where the loops are absent and where the nonlinear dispersion has two bands for  fixed quasimomentum [depicted  by thin lines in the left panel of  Fig.~\ref{Quench}(b)]. The quench dynamics is again periodic [see the right panel], but its oscillation frequency still does not correspond to the energy difference between the two bands.

If one imagines the quench of the sign of $\Omega$ as a perturbation to the ground states, one can have a collective excitation dispersion relations $\omega(q)$ with only two bands. These collective excitation bands are shown in the left panels by thick green lines. They are calculated from the Bogoliubov de Gennes equations~\cite{Zheng2013, Martone}, and in experiments can be measured by Bragg spectroscopy~\cite{Khamehchi2014, Pan2015}. We have checked that the period of the quench dynamics has no relation with the energy difference between the two collective excitation bands.

The quench dynamics experiment described by Qu {\it et al.} in~\cite{Qu} surprisingly connects with the single-particle dispersion, as in their $^{87}$Rb atomic BECs, the nonlinear coefficients approximately satisfy $g_{12}=g$. According to our nonlinear two-level model in Eq.~(\ref{NTL}), one sees that their system is effectively a single-particle model, since the nonlinear term disappears when $g_{12}-g=0$.
For the case of  $g_{12}-g=0$, nonlinear dispersion is just an overall displacement of linear dispersion, which can be seen from Eq.~(\ref{NTL}) and $\mu= \mu_\text{ntl} + (g+g_{12})/2 + k^2/2$. In Fig.~\ref{Quench}(c), we show 	the nonlinear dispersion for  $g_{12}=g$ and verify that the oscillation frequency of the quench dynamics in the right panel matches up with the energy difference between two circles in the left. It is very interesting to note that the oscillation frequency also equals  the difference between two collective excitation bands labeled by thick green lines. The quench dynamics is interpreted as Zitterbewengung due to the superposition of two bands in the experiment~\cite{Qu}. With the nonlinearity which breaks quantum superposition, it is ambiguous to simulate Zitterbewengung in the current system.

As there is no direct relation between the quench dynamics and the nonlinear dispersion relation, the quench dynamics may be understood from a semiclassical analysis~\cite{Smerzi}. By defining $\Phi_1=|\Phi_1|e^{i\theta_1}$, $\Phi_2=|\Phi_2|e^{i\theta_2}$, and $n=|\Phi_2|^2-|\Phi_1|^2$, $\theta= \theta_2- \theta_1$, the nonlinear two-level model in Eq.~(\ref{NTL}) has a classical Hamiltonian as $H_c= (g_{12}-g)n^2/2 +2(\gamma k +\delta/2)n -\Omega \sqrt{1-n^2} \cos(\theta)$~\cite{Liu}. In the phase space of ($n, \theta$), the oscillations induced by the quench are closed orbits around the fixed points of $H_c$, which feature one period. The oscillation frequency depends on the nonlinearity~\cite{Liu, Smerzi}.

In conclusion, we have investigated two kinds of nonlinear dynamics in spin-orbit-coupled BECs. For the adiabatic motions induced by a weak constant acceleration force, the atoms undergo nonlinear Landau-Zener tunneling in the presence of  interesting loop structures in the nonlinear dispersion relation. At the same time, for a quench dynamics we have found that no connection to the nonlinear dispersion relation exists.


 We sincerely acknowlege Thomas Busch for helpful discussion and revising the manuscript, and we are grateful to Peter Engels, Qiongtao Xie, Biao Wu, Chuanwei Zhang, and Chunlei Qu for very constructive discussions. YZ acknowledges the hospitality of Okinawa Institute of Science and Technology Graduate University, where part of the study was initiated. This work was supported by the NSF of China (Grant No. 11774219), by the Thousand Young Talents Program of China, and by the Eastern Scholar and Shuguang (17SG39) Program of Shanghai.

\end{document}